\documentstyle[buckow]{article}

\newcommand{\be}{\begin{equation}}
\newcommand{\ee}{\end{equation}}
\newcommand{\nn}{\nonumber}
\newcommand{\bea}{\begin{eqnarray}}
\newcommand{\eea}{\end{eqnarray}}

\newcommand{\tr}{\textrm{tr}}
\newcommand{\ns}{\normalsize}

\def\b{\beta}
\def\g{\gamma}

\def\d{\delta}
\def\e{\epsilon}

\def\k{\kappa}

\def\m{\mu}
\def\n{\nu}
\def\o{\omega}
\def\p{\pi}

\def\r{\rho}

\def\cB{{\cal B}}

\def\cN{{\cal N}}

\def\cC{{\cal C}}

\def\Ib{\bar{I}}
\def\Jb{\bar{J}}
\def\Kb{\bar{K}}
\def\Lb{\bar{L}}

\def\bbar{\bar{b}}

\begin{document}
\hfill{\ns OUTP-99-20P, hep-th/9903144}\\
\def\titleline{Heterotic M--Theory Vacua with Five--Branes
 \footnote{Talk presented by A.~Lukas at the 32nd Symposium
 Ahrenshoop on the Theory of Elementary Particles,
Buckow, Germany, September 1 - 5, 1998}
}
\def\authors{Andr\'e Lukas\1ad, Burt A.~Ovrut\2ad and Daniel Waldram\3ad}
\def\addresses{
\1ad{Department of Physics, Theoretical Physics, 
     University of Oxford \\
     1 Keble Road, Oxford OX1 3NP, United Kingdom} \\[0.2cm]
\2ad{Department of Physics, University of Pennsylvania\\
     Philadelphia, PA 19104--6396, USA} \\[0.2cm]
\3ad{Department of Physics, Joseph Henry Laboratories, Princeton University\\
     Princeton, NJ 08544, USA}}
\def\abstracttext{We construct vacua of heterotic M--theory with
     general gauge bundles and five--branes. Some aspects of the
     resulting low--energy effective theories are discussed.}
\large
\makefront
 
M--theory on the orbifold $S^1/Z_2$ is believed to describe the
strong coupling limit of the $E_8\times E_8$ heterotic
string~\cite{hw1} and, therefore, constitutes a particularly interesting
starting point for M--theory particle phenomenology. At low energy,
this theory is described by 11--dimensional supergravity coupled to
two 10--dimensional $E_8$ gauge multiplets residing on the two fixed
points of the orbifold~\cite{hw2,hor,w}. Compactifications leading to
$\cN =1$ supersymmetry in four dimensions are based on space--times
of the structure
\begin{equation} 
M_{11}= S^1/Z_2\times X\times M_4,\label{st}
\end{equation}
where $X$ is a Calabi--Yau three-fold and $M_4$ is flat Minkowski space.
While most work to date, related to such compactifications of heterotic
M--theory, has been limited to the standard embedding,
non--standard embedding vacua have previously been addressed
in~\cite{benakli,stieb,lpt}. 

Here, we will consider the general configuration leading to $\cN=1$
supersymmetry~\cite{nse}, where, first, we allow for general gauge
bundles, and, second, include five-branes~\cite{w}, states which
are essentially non-perturbative in heterotic string theory.
The important new ingredient is the presence of these five--branes in
the vacua. Some effects caused by the presence of the five--branes
have been discussed in~\cite{stieb,BDDR}. Due to gauge and gravity
sources on the orbifold planes
and the five--brane sources, the space--time~(\ref{st}) receives
corrections that can be computed perturbatively as an expansion in~\cite{bd}
\begin{equation}
 \e_S = \left(\frac{\k}{4\p}\right)^{2/3}
        \frac{2\p\r}{v^{2/3}},\label{es}
\end{equation}
where $\k$, $v$ and $\r$ are the 11--dimensional Newton constant, the
Calabi--Yau volume and the orbifold radius, respectively. 

\vspace{0.4cm}

Let us now determine these corrections to linear order in $e_S$.
We need to specify the 11--dimensional metric $g_{IJ}$, the
three--form $C_{IJK}$ with field strength
$G_{IJKL}=24\,\partial_{[I}C_{JKL]}$ (where $I,J,K,\dots = 0,\dots
,9,11$) and the Killing spinor $\eta$. This can be done solving the
Bianchi identity
\bea
 (dG)_{11\Ib\Jb\Kb\Lb} &=& 4\sqrt{2}\p\left(\frac{\k}{4\p}
                           \right)^{2/3}\left[J^{(0)}\d (x^{11})+J^{(N+1)}
                           \d (x^{11}-\p\r )+\right.\nn \\
                       &&\left.\qquad\qquad\qquad\qquad\frac{1}{2}
                         \sum_{n=1}^NJ^{(n)}(\d (x^{11}-x_n)+\d (x^{11}+x_n))
                           \right]_{\Ib\Jb\Kb\Lb}\; .\label{G}
\eea
where $\Ib ,\Jb ,\Kb ,\dots = 0,\dots ,9$, along with the equation of
motion for $C$ and the Killing spinor equation $\d\Psi_I =0$ for
the gravitino to preserve some supersymmetry. The sources on the
orbifold planes
\begin{equation}
J^{(0)} = -\frac{1}{16\p^2}\left(\tr {F^{(1)}}^2 
- \frac{1}{2}\tr R^2\right)\; ,\qquad
J^{(N+1)} = -\frac{1}{16\p^2}\left(\tr {F^{(2)}}^2 
- \frac{1}{2}\tr R^2\right)
\end{equation}
are given in term of the $E_8$ gauge field strengths $F^{(i)}$,
$i=1,2$, and the curvature. In addition, we have considered $N$ five--branes
transverse to the orbifold. They induce the sources
$J^{(n)}$, $n=1,\dots ,N$ appearing in the above Bianchi
identity. We have chosen the orbifold coordinate $x^{11}$ to be in
the range $x^{11}\in [-\p\r ,\p\r ]$ with the orbifold planes at
$x^{11}=0,\p\r$ and the five--branes at $x^{11}=\pm x_1,\dots ,\pm
x_N$. From eq.~(\ref{G}), the sum of all these sources must be
cohomologically trivial.

One must also ensure that the theories on the orbifold planes and the
five--branes preserve supersymmetry. This is guaranteed by choosing
two semi--stable holomorphic (otherwise arbitrary) $E_8$ gauge bundles
$V_i$ and by wrapping the five--branes on holomorphic Calabi--Yau
two--cycles while stretching them over the four--dimensional
uncompactified space. With this choice, all sources $J^{(n)}$,
$n=0,\dots ,N+1$ are $(2,2)$ forms on the Calabi--Yau space. The
cohomology condition on the sources can now be expressed as 
\begin{equation}
   \left[\sum_{n=0}^{N+1}J^{(n)}\right] =
   c_2(V_1)+c_2(V_2)-c_2(TX)+[W] = 0\; .\label{coh}
\end{equation}
Here $c_2(V_i)$ and $c_2(TX)$ are the second Chern classes of the
vector bundles $V_i$ and the tangent bundle $TX$ and
$[W]=\sum_{n=1}^N[J^{(n)}]$ is the cohomology class of the total
five--brane curve $W$. We write
\begin{equation}
    g_{IJ} = g_{IJ}^{(0)}+g_{IJ}^{(1)}\; ,\qquad
    G_{IJKL} = G_{IJKL}^{(1)}\; ,\qquad
    \eta = \eta^{(0)}+\eta^{(1)}\; ,
 \label{decomp1}
\end{equation}
where $g_{IJ}^{(0)}$ and $\eta^{(0)}$ are the metric and the
covariantly constant spinor of the Calabi--Yau space $X$. One can
show~\cite{w} that $G^{(1)}_{ABCD}$ and $G^{(1)}_{ABC11}$ are the only
non--vanishing components of the four--form and that the other
corrections have the structure
\begin{equation}
 g^{(1)}_{\m\n} = b\eta_{\m\n}\; ,\qquad g^{(1)}_{AB} = h_{AB}\; ,\qquad
 g^{(1)}_{11,11} = \g\; ,\qquad \eta^{(1)} = \psi\eta^{(0)}\; .
\end{equation}
Four--dimensional space is indexed by $\m ,\n ,\r\dots = 0,\dots ,3$
while we use $A,B,C,\dots = 4,\dots ,9$ for the Calabi--Yau space.
Furthermore, holomorphic (anti--holomorphic) Calabi--Yau indices are
denoted by $a,b,c, \dots$ ($\bar{a} ,\bar{b} ,\bar{c} ,\dots$). The
corrections can be entirely expressed in terms of a $(1,1)$ form
$\cB_{a\bar{b}}$ on the Calabi--Yau space as~\cite{low}
\begin{equation}
\begin{array}{rlllrlll}
  h_{a\bbar} &=& \sqrt{2}i \left( \cB_{a\bbar} 
     - \frac{1}{3}\o_{a\bbar}\cB \right) &,&
  b &=& \frac{\sqrt{2}}{6} \cB  \\
  \g &=& -\frac{\sqrt{2}}{3} \cB &,&
  \psi &=& -\frac{\sqrt{2}}{24} \cB  \\
  G_{ABCD}^{(1)} &=& \frac{1}{2}\e_{ABCDEF}\partial_{11}\cB^{EF} &,&
  G_{ABC11}^{(1)} &=& \frac{1}{2}\e_{ABCDEF}\partial^D\cB^{EF} 
\end{array}
\label{sol} 
\end{equation}
where $\cB = \o^{AB}\cB_{AB}$ and $\o_{a\bar{b}}=-ig_{a\bar{b}}$ is the 
K\"ahler form. One can expand this $(1,1)$ form in terms of
eigenfunctions of the Calabi--Yau Laplacian as
\begin{equation}
 \cB_{AB} = \sum_ib_i\o_{AB}^i+\mbox{massive}\; .\label{bser}
\end{equation}
Here, we have concentrated on the massless part of this expansion, 
represented by the harmonic $(1,1)$ forms $\o_{ia\bar{b}}$ of the
Calabi--Yau space, where $i=1,\dots ,h^{1,1}$. The form of
the massive part can be found in \cite{nse}. The expansion
coefficients $b_i$ read explicitly
\begin{equation}
 b_{i} = \frac{\p\e_S}{\sqrt{2}}\left[\sum_{m=0}^n
       \b_{i}^{(m)}(|z|-z_m)-\frac{1}{2}\sum_{m=0}^{N+1}(z_m^2-2z_m)
       \b_{i}^{(m)}\right]\; ,\qquad 
 \b_{i}^{(n)} = \int_{\cC_{4i}}J^{(n)}\; .
\label{massless}
\end{equation}
This expression holds in each interval $z_n\leq |z|\leq z_{n+1}$ for
fixed $n$, where $n=0,\dots ,N$. Here $z=x^{11}/\p\r$ and
$z_n=x_n/\p\r$ are normalized orbifold coordinates
and $\cC_{4i}$ is a basis of four--cycles dual to the harmonic
$(1,1)$ forms $\o_i$. Note that the charges sum up to zero due to the
cohomology constraint (\ref{coh}); that is $\sum_{n=0}^{N+1}\b_{i}^{(n)} = 0$.

To summarize, eqs.~(\ref{decomp1})--(\ref{massless}) represent the
(massless part) of the background to linear order in $\e_S$, given in
terms of the topological charges $\b_{i}^{(n)}$. Specific consistent
models where these charges can be computed explicitly have been constructed
recently~\cite{np}. Eq.~(\ref{massless}) shows that the Calabi--Yau
space is changing its size and shape across the orbifold. This
corresponds to a certain trajectory in the K\"ahler moduli space.

\vspace{0.4cm}

We would now like to discuss some properties of the low--energy
effective actions associated to those vacua. Interesting new features
arise from the new degrees of freedom on the five--brane
worldvolumes. From a single five--brane wrapped on a holomorphic
genus $g$ two--cycle within a Calabi--Yau space one obtains an
$\cN =1$ theory  in four dimensions with generically $g$ $U(1)$ vector
multiplets, one universal chiral multiplet (containing the position
modulus in the orbifold direction) and a number of additional chiral
multiplets parameterizing the moduli space of the curve. In specific
cases, the gauge group enhances and becomes non--abelian. For example,
$N$ five--branes, wrapped on the same cycle with genus $g$ and positioned at
different points in the orbifold lead to a group $U(1)^{gN}$. Moving
all five--branes to the same orbifold point enhances the group to
$U(N)^g$. 

The five--dimensional effective action of heterotic M--theory obtained
by reducing on the Calabi--Yau space is of some interest as the
orbifold size could be large. For standard embedding, this action
consists of a gauged $\cN =1$ supergravity with vector and
hypermultiplets in the bulk coupled to two four--dimensional $\cN =1$
theories on the orbifold planes~\cite{losw}. For the vacua discussed
here, the bulk action between each two neighboring five--branes is
again given by gauged supergravity. The gauge charges are directly
related to the charges $\b_i^{(n)}$ and depend
on the five--brane pair considered. In addition to the two orbifold
theories, the bulk is now coupled to $N$ additional four--dimensional
$\cN =1$ theories resulting from the five--branes. Each of them
carries a field content of the type discussed above.

Upon further reduction to four dimensions, one obtains a theory which
consists of the usual ``observable'' and ``hidden'' sector originating
from $E_8\times E_8$ and $N$ additional sectors from the five--brane
degrees of freedom. In particular, the low--energy gauge group is
enhanced to $G^{(1)}\times G^{(2)}\times G$ where $G^{(i)}$ are the
unbroken subgroups of $E_8$ and $G$, being typically a product of
unitary groups, results from the five branes. In addition to the existence
of new sectors, the ``conventional'' $E_8\times E_8$ sectors are
effected by the presence of the five--branes. The five--brane charges
introduce more freedom in the cohomology condition~(\ref{coh}) and,
hence, allow for consistent gauge bundles that would otherwise be
forbidden. Among other things, this facilitates the construction of
three--family model~\cite{np}. Also, the conventional part of the
four--dimensional effective action receives new corrections caused
by the five--branes. They can be computed in leading order using the
above results for the explicit form of the vacua. For example, the
gauge kinetic functions for $G^{(1)}$ and $G^{(2)}$ take the form
\begin{equation}
 f^{(1)} = S + \p\e_S T^{i}\sum_{n=0}^{N+1}(1-z_n)^2
             \b_{i}^{(n)} \; ,\qquad
 f^{(2)} = S + \p\e_S T^{i}\sum_{n=1}^{N+1}{z_n}^2
             \b_{i}^{(n)} \; ,
\end{equation}
where $S$ is the dilaton and $T^i$ are the $T$ moduli. Note the
dependence on the five--brane charges as well as on the positions
$z_n$ of the five--branes in the orbifold.

\vskip0.5cm
\noindent
{\large \bf Acknowledgments}

\smallskip
\noindent
A.~L.~is supported by the European Community
under contract No.~FMRXCT 960090. B.~A.~O.~is supported in part by 
DOE under contract No.~DE-AC02-76-ER-03071 and by a Senior 
Alexander von Humboldt Award. D.~W.~is supported in part by
DOE under contract No.~DE-FG02-91ER40671.

%%%%%%%%%%%%%%%%%%%%%%
%%%%%%%%%%%%%%%%%%%%%%

\end{document}